\documentclass[12pt,titlepage,letterpaper]{utarticle}

\usepackage{amsmath}
\usepackage{mathrsfs}
\usepackage{amsfonts}
\usepackage{amssymb}
\usepackage{amsthm}
\usepackage{mathtools}
\usepackage{mathbbol}
\usepackage{graphicx}
\usepackage{color}
\usepackage{xparse}
\usepackage{tikz}
\usepackage{tocloft}
\usepackage[titletoc,title]{appendix}
\usepackage[nosort]{cite}
\usepackage{hyperref}
\usepackage{longtable}
\usepackage{comment}

\usepackage{caption}

\usetikzlibrary{cd}

\definecolor{lightmauve}{RGB}{255,187,255}
\definecolor{lightblue}{RGB}{238,238,255}
\definecolor{lightred}{RGB}{255,238,238}
\definecolor{midgreen}{RGB}{15,170,15}

\newcommand{\DEnext}{\cite{Distler:2024whoknows}}

\definecolor{aqua}{rgb}{0, 1.0, 1.0}
\definecolor{fuschia}{rgb}{1.0, 0, 1.0}
\definecolor{gray}{rgb}{0.502, 0.502, 0.502}
\definecolor{lime}{rgb}{0, 1.0, 0}
\definecolor{maroon}{rgb}{0.502, 0, 0}
\definecolor{navy}{rgb}{0, 0, 0.502}
\definecolor{olive}{rgb}{0.502, 0.502, 0}
\definecolor{purple}{rgb}{0.502, 0, 0.502}
\definecolor{silver}{rgb}{0.753, 0.753, 0.753}
\definecolor{teal}{rgb}{0, 0.502, 0.502}

\makeatletter
\newdimen\itex@wd%
\newdimen\itex@dp%
\newdimen\itex@thd%
\def\itexspace#1#2#3{\itex@wd=#3em%
\itex@wd=0.1\itex@wd%
\itex@dp=#2ex%
\itex@dp=0.1\itex@dp%
\itex@thd=#1ex%
\itex@thd=0.1\itex@thd%
\advance\itex@thd\the\itex@dp%
\makebox[\the\itex@wd]{\rule[-\the\itex@dp]{0cm}{\the\itex@thd}}}
\makeatother

\makeatletter
\newif\if@sup
\newtoks\@sups
\def\append@sup#1{\edef\act{\noexpand\@sups={\the\@sups #1}}\act}%
\def\reset@sup{\@supfalse\@sups={}}%
\def\mk@scripts#1#2{\if #2/ \if@sup ^{\the\@sups}\fi \else%
  \ifx #1_ \if@sup ^{\the\@sups}\reset@sup \fi {}_{#2}%
  \else \append@sup#2 \@suptrue \fi%
  \expandafter\mk@scripts\fi}
\def\tensor#1#2{\reset@sup#1\mk@scripts#2_/}
\def\multiscripts#1#2#3{\reset@sup{}\mk@scripts#1_/#2%
  \reset@sup\mk@scripts#3_/}
\makeatother

\makeatletter
\newbox\slashbox \setbox\slashbox=\hbox{$/$}
\def\itex@pslash#1{\setbox\@tempboxa=\hbox{$#1$}
  \@tempdima=0.5\wd\slashbox \advance\@tempdima 0.5\wd\@tempboxa
  \copy\slashbox \kern-\@tempdima \box\@tempboxa}
\def\slash{\protect\itex@pslash}
\makeatother

\def\clap#1{\hbox to 0pt{\hss#1\hss}}

\let\oldroot\root
\def\root#1#2{\oldroot #1 \of{#2}}
\renewcommand{\sqrt}[2][]{\oldroot #1 \of{#2}}

\DeclareSymbolFont{symbolsC}{U}{txsyc}{m}{n}
\SetSymbolFont{symbolsC}{bold}{U}{txsyc}{bx}{n}
\DeclareFontSubstitution{U}{txsyc}{m}{n}

\DeclareSymbolFont{stmry}{U}{stmry}{m}{n}
\SetSymbolFont{stmry}{bold}{U}{stmry}{b}{n}

\DeclareFontFamily{OMX}{MnSymbolE}{}
\DeclareSymbolFont{mnomx}{OMX}{MnSymbolE}{m}{n}
\SetSymbolFont{mnomx}{bold}{OMX}{MnSymbolE}{b}{n}
\DeclareFontShape{OMX}{MnSymbolE}{m}{n}{
    <-6>  MnSymbolE5
   <6-7>  MnSymbolE6
   <7-8>  MnSymbolE7
   <8-9>  MnSymbolE8
   <9-10> MnSymbolE9
  <10-12> MnSymbolE10
  <12->   MnSymbolE12}{}

\makeatletter
\def\re@DeclareMathSymbol#1#2#3#4{%
    \let#1=\undefined
    \DeclareMathSymbol{#1}{#2}{#3}{#4}}
\re@DeclareMathSymbol{\neArrow}{\mathrel}{symbolsC}{116}
\re@DeclareMathSymbol{\neArr}{\mathrel}{symbolsC}{116}
\re@DeclareMathSymbol{\seArrow}{\mathrel}{symbolsC}{117}
\re@DeclareMathSymbol{\seArr}{\mathrel}{symbolsC}{117}
\re@DeclareMathSymbol{\nwArrow}{\mathrel}{symbolsC}{118}
\re@DeclareMathSymbol{\nwArr}{\mathrel}{symbolsC}{118}
\re@DeclareMathSymbol{\swArrow}{\mathrel}{symbolsC}{119}
\re@DeclareMathSymbol{\swArr}{\mathrel}{symbolsC}{119}
\re@DeclareMathSymbol{\nequiv}{\mathrel}{symbolsC}{46}
\re@DeclareMathSymbol{\Perp}{\mathrel}{symbolsC}{121}
\re@DeclareMathSymbol{\Vbar}{\mathrel}{symbolsC}{121}
\re@DeclareMathSymbol{\sslash}{\mathrel}{stmry}{12}
\re@DeclareMathSymbol{\boxslash}{\mathrel}{stmry}{27}
\re@DeclareMathSymbol{\boxbslash}{\mathrel}{stmry}{28}
\re@DeclareMathSymbol{\boxast}{\mathrel}{stmry}{24}
\re@DeclareMathSymbol{\boxcircle}{\mathrel}{stmry}{29}
\re@DeclareMathSymbol{\boxbox}{\mathrel}{stmry}{30}
\re@DeclareMathSymbol{\obslash}{\mathrel}{stmry}{20}
\re@DeclareMathSymbol{\obar}{\mathrel}{stmry}{58}
\re@DeclareMathSymbol{\olessthan}{\mathrel}{stmry}{60}
\re@DeclareMathSymbol{\ogreaterthan}{\mathrel}{stmry}{61}
\re@DeclareMathSymbol{\bigsqcap}{\mathop}{stmry}{"64}
\re@DeclareMathSymbol{\biginterleave}{\mathop}{stmry}{"6}
\re@DeclareMathSymbol{\invamp}{\mathrel}{symbolsC}{77}
\re@DeclareMathSymbol{\parr}{\mathrel}{symbolsC}{77}
\makeatother

\makeatletter
\def\Decl@Mn@Delim#1#2#3#4{%
  \if\relax\noexpand#1%
    \let#1\undefined
  \fi
  \DeclareMathDelimiter{#1}{#2}{#3}{#4}{#3}{#4}}
\def\Decl@Mn@Open#1#2#3{\Decl@Mn@Delim{#1}{\mathopen}{#2}{#3}}
\def\Decl@Mn@Close#1#2#3{\Decl@Mn@Delim{#1}{\mathclose}{#2}{#3}}
\Decl@Mn@Open{\llangle}{mnomx}{'164}
\Decl@Mn@Close{\rrangle}{mnomx}{'171}
\Decl@Mn@Open{\lmoustache}{mnomx}{'245}
\Decl@Mn@Close{\rmoustache}{mnomx}{'244}
\Decl@Mn@Open{\llbracket}{stmry}{'112}
\Decl@Mn@Close{\rrbracket}{stmry}{'113}
\makeatother

\makeatletter
\DeclareRobustCommand\widecheck[1]{{\mathpalette\@widecheck{#1}}}
\def\@widecheck#1#2{%
    \setbox\z@\hbox{\m@th$#1#2$}%
    \setbox\tw@\hbox{\m@th$#1%
       \widehat{%
          \vrule\@width\z@\@height\ht\z@
          \vrule\@height\z@\@width\wd\z@}$}%
    \dp\tw@-\ht\z@
    \@tempdima\ht\z@ \advance\@tempdima2\ht\tw@ \divide\@tempdima\thr@@
    \setbox\tw@\hbox{%
       \raise\@tempdima\hbox{\scalebox{1}[-1]{\lower\@tempdima\box
\tw@}}}%
    {\ooalign{\box\tw@ \cr \box\z@}}}
\makeatother

\makeatletter
\NewDocumentCommand\mathraisebox{moom}{%
\IfNoValueTF{#2}{\def\@temp##1##2{\raisebox{#1}{$\m@th##1##2$}}}{%
\IfNoValueTF{#3}{\def\@temp##1##2{\raisebox{#1}[#2]{$\m@th##1##2$}}%
}{\def\@temp##1##2{\raisebox{#1}[#2][#3]{$\m@th##1##2$}}}}%
\mathpalette\@temp{#4}}
\makeatletter

\makeatletter
\def\udots{\mathinner{\mkern2mu\raise\p@\hbox{.}
\mkern2mu\raise4\p@\hbox{.}\mkern1mu
\raise7\p@\vbox{\kern7\p@\hbox{.}}\mkern1mu}}
\makeatother

\theoremstyle{plain}

\theoremstyle{definition}

\theoremstyle{remark}

\def\Bid{{\mathchoice {\rm {1\mskip-4.5mu l}} {\rm
{1\mskip-4.5mu l}} {\rm {1\mskip-3.8mu l}} {\rm {1\mskip-4.3mu l}}}}

\usetikzlibrary{positioning,positioning,arrows,decorations.markings}

\begin{document}
\renewcommand{\arraystretch}{1.5}

%-------------------------------------------------------------------
\preprint{
UTWI--39--2024\\
}

\title{Even More On Twisted $A_{2n}$ Class-S Theories}

\author{Jacques Distler, Grant Elliot
     \oneaddress{
      Weinberg Institute for Theoretical Physics\\
      Department of Physics,\\
      University of Texas at Austin,\\
      Austin, TX 78712, USA \\
      {~}\\
      \email{distler@golem.ph.utexas.edu}\\
      \email{gelliot123@utexas.edu}
      }
}
\date{\today}
%\date{August 30, 2023}

\Abstract{
This paper is a continuation of our investigation into the Coulomb branches of twisted $A_{2n}$ of class-S. In \DEnext, we found predictions for the contributions of twisted punctures to the graded dimensions of the Coulomb branch, based on the behaviour under nilpotent Higgsings and S-duality. While surprisingly powerful, these arguments were indirect. Here, we take a different approach: we define precisely the nature of the automorphism under which the twisted punctures are twisted (in particular, it is order-4, not order-2). From that, we find the local constraints satisfied by the Laurent coefficients of the invariant polynomials in the Higgs field, for all twisted punctures in $A_{2n}$, for all $n$. A crucial role is played by a new (at least, new in physics) order-reversing map on the set of nilpotent orbits in $\mathfrak{sp}(n)$. Finally, we construct several examples of Seiberg-Witten curves  for 3-punctured spheres in these theories. 
}

\maketitle

\tocloftpagestyle{empty}
\tableofcontents
\vfill
\newpage
\setcounter{page}{1}

\section{Introduction}

The study of four dimensional theories with $\mathcal{N}=2$ supersymmetry was revolutionized with the discovery that the low energy effective field theory on the Coulomb branch is determined by a complex integrable system\cite{Seiberg:1994aj}. Frequently, this turned out to be Hitchin's integrable system. The explanation for the ubiquity of Hitchin systems in 4d $\mathcal{N}=2$ came with the class-S construction \cite{Gaiotto:2009we,Gaiotto:2009hg}, where the 4d theory is obtained by compactifying the 6d $\mathcal{N}=(2,0)$ theory on a Riemann surface $C$. The Coulomb branch geometry is encoded in a Hitchin system on $C$. Codimension-two defects can be placed at points (punctures) on $C$ and result in singularities for the Higgs fields at said points. 

When the $(2,0)$ theory has a discrete zero-form symmetry, the class-S construction can be supplemented by the presence of twist lines, which correspond to some nontrivial background for the symmetry along the Riemann surface. Twist lines can emanate from a codimension-two defect, and the corresponding boundary conditions for the Higgs field were determined in \cite{Chacaltana:2012zy}, except for twisted defects in the $A_{2n}$ case.

This particular case was first recognized to be extraordinary in \cite{Tachikawa:2011ch}. The central issue with these theories is that the singular behavior for the Hitchin system due to the codimension-two defects was unknown. However, many other aspects of these theories have been studied such as their superconformal indices\cite{Gadde:2011ik,Mekareeya:2012tn}, VOAs\cite{Beem:2013sza,Beem:2014rza,Arakawa:2018egx,Beem:2022mde,Elliot:2024hat}, 3d Mirrors\cite{Benini:2010uu,Beratto:2020wmn,Kang:2022zsl}, and the global anomaly associated to twisted punctures\cite{Tachikawa:2018rgw}. Utilizing what was known about these theories, \DEnext\ determined the precise singular behavior of the Higgs field at twisted punctures. 

In this paper, we determine the boundary conditions satisfied by the Higgs field at the punctures, and thus are able to construct corresponding Hitchin systems or, at least, write down the spectral (or Seiberg-Witten) curve. The invariant polynomials in the Higgs field are holomorphic $k$-differentials with specified singularities at the punctures (determined by a choice of nilpotent orbit for the residue of the Higgs field at each puncture). Naively, the base of the Hitchin integrable system is the vector space of these holomorphic $k$-differentials. However, that's not quite right. The coefficients of the Laurent expansions of the $k$-differentials in the neighbourhood of the puncture are not independent. They obey polynomial constraints and the ``true'' Hitchin base is obtained by solving those constraints.

The (graded) dimension of the space of solutions to those constraints agrees with the previous predictions of \DEnext\ and \cite{Beem:2020pry}.  But here, we can do more and actually construct the Seiberg-Witten curves. We do so for a number of examples.

\section{Local Behaviour}

\subsection{Outer Automorphisms}
Before considering the $A_{2n}$ case, let us recall the outer-automorphism acting in the twisted $A_{2n-1}$ case. The automorphism acts as $x \to -Rx^tR^{-1}$, where $R$ is given by $i \sigma_2 \otimes \Bid_{n}$. If we take $x$ to be a block matrix, consisting of $n\times n$ blocks (with the restriction $tr(A)=-tr(D)$), the automorphism acts as 
\begin{equation*}
    \left(
\begin{array}{cc}
 A & B\\
  C & D  \\
\end{array}
\right) \to \left(
\begin{array}{cc}
 -D^t & B^t\\
  C^t & -A^t  \\
\end{array}
\right).
\end{equation*}
The invariant subalgebra is precisely $Sp(n)$.

In the $A_{2n}$ case, we consider the outer-automorphism $\alpha: x \to -Rx^tR^{-1}$, where $R$ is given by $(1\oplus(i \sigma_2 \otimes \Bid_{n}))$. More concretely, in block notation (where the $v_i$ and $w_i$ are $n$-component column vectors)
\[
\alpha: \begin{pmatrix}
-tr(A)-tr(D)&v_1^t&v_2^t\\
w_1& A&B\\
w_2&C&D
\end{pmatrix}
\to
\begin{pmatrix}
tr(A)+tr(D)&w_2^t&-w_1^t\\
v_2& -D^t&B^t\\
-v_1&C^t&-A^t
\end{pmatrix}
\]
$\alpha$ has order four as $R^2=1\oplus(-\Bid_{2n})$. The invariant subalgebra is
\[
\mathfrak{j}_+ =\left\{\left.
\left(\begin{smallmatrix}
0&0&0\\
0&A&B\\
0&C&-A^t
\end{smallmatrix}\right)\right\vert A\;\text{arbitrary}, B^t=B, C^t=C
\right\} =\mathfrak{sp}(n)
\]
and the other eigenspaces are
\[
\begin{split}
\mathfrak{j}_{ i}&=
\left\{
\left(
\begin{smallmatrix}
0&v_1^t&v_2^t\\
-iv_2&0&0\\
iv_1&0&0
\end{smallmatrix}
\right)
\right\}= 
\begin{tikzpicture} 
 \draw[black,fill=white] (0.25, -0.25) rectangle (0.5, -0.5);
\end{tikzpicture} \\
\mathfrak{j}_{-1}&=
\left\{\left.
\left(
\begin{smallmatrix}
-2tr(D)&0&0\\
0&D&E\\
0&F&D^t
\end{smallmatrix}
\right)
\right\vert D\;\text{arbitrary}, E^t=-E, F^t=-F
\right\} =1+
\begin{tikzpicture} 
 \draw[black,fill=white] (0.25, -0.25) rectangle (0.5, -0.5);
  \draw[black,fill=white] (0.25, -0.5) rectangle (0.5, -0.75);
\end{tikzpicture} \\
\mathfrak{j}_{- i}&=
\left\{
\left(
\begin{smallmatrix}
0&w_1^t&w_2^t\\
iw_2&0&0\\
-iw_1&0&0
\end{smallmatrix}
\right)
\right\}= 
\begin{tikzpicture} 
 \draw[black,fill=white] (0.25, -0.25) rectangle (0.5, -0.5);
\end{tikzpicture} 
%\begin{tikzpicture} 
%\begin{scope}
%      \foreach \p [count = \r] in {1,1}{
%        \foreach \i in {1,...,\p}{
%          \draw[black,fill=white] (0.25*\i, -0.25*\r) rectangle (0.25*\i+0.25, -0.25-0.25*\r);
%        }
%      }
%  \end{scope}
%\end{tikzpicture} 
\end{split}
\]
The $A_{2n}$ algebra decomposes under this embedding of $\mathfrak{sp}(n)$ as $\mathfrak{sp}(n)$ plus two fundamentals plus a traceless-antisymmetric plus a singlet.

\subsection{The Hitchin Image}

 The Coulomb branches of class-S theories obtained from compactifying a type $\mathfrak{j}$ $\mathcal{N}=(2,0)$ theory are Hitchin systems of type $\mathfrak{j}$. One may add twist lines corresponding to a background field for the discrete global symmetry of the 6d theory, which implement a monodromy when going around various cycles of the Riemann surface. Additionally, codimension-two defects can be placed at points of the Riemann surface, which result in singularities for the Higgs field. For regular defects, the Higgs field only has simple poles. Let $\mathfrak{g}^{\vee}$ be the subalgebra invariant under the monodromy around the singularity, and denote its Langlands dual as $\mathfrak{g}$. The residue of the Higgs field at a twisted puncture is a nilpotent element of $\mathfrak{g}^{\vee}$. However, not all nilpotent elements are allowed. In the familiar class-S case, such orbits are restricted to those that are special, that is they are in the image of the Spaltenstein map. A twisted regular codimension-two defect is labeled by a nilpotent orbit of $\mathfrak{g}$. Its image under the Spaltenstein(-Barbasch-Vogan) map $d:\text{Nilp}(\mathfrak{g})\to \text{Nilp}(\mathfrak{g}^\vee)$, called the Hitchin orbit, determines the residue at the simple pole of the Higgs field. Thus, only special orbits are seen in the Hitchin systems for ordinary class-S theories.

In the twisted $A_{2n}$ case, the Nahm orbits are labeled by $C_{n}$ orbits, not $B_{n}$ orbits. Thus, there should be some order reversing map from $C_n$ orbits to $C_{n}$ orbits determining the Hitchin partition. It was argued in \DEnext\ that this map should be the one found in \cite{MR1404331}, denoted $d':\text{Nilp}(\mathfrak{c}_n)\to \text{Nilp}(\mathfrak{c}_n)$. The orbits in the image of this map are called metaplectic special (or anti-special).  $d'$ can be given the following algorithmic description\cite{MR3666056}: Append 1 to the partition, then take the transpose and $C$-collapse. This will not actually be a $C$-partition, so we must subtract one from the last part of the partition. 

We now briefly summarize the reasoning behind this map. In other class-S theories, punctures in the same special piece have the same Hitchin orbit, but the choice of Nahm orbit effects the choice of $a$-type constraints. Thus, replacing a puncture with another in the same special piece does not change the dimension of the Coulomb branch. For classical Lie algebras, this can be seen through nilpotent Higgsing of $Sp(n)$ factors at odd level \cite{Distler:2022nsn,Distler:2022jvk}. These odd level factors necessarily possess Witten's global anomaly \cite{Witten:1982fp}, which results in the theory having an extended Coulomb branch. Contrarily, in the twisted $A_{2n}$ case, $Sp(n)$ symmetries at even level will have Witten's global anomaly \cite{Tachikawa:2018rgw}. Hence, we expect nilpotent Higgsings of even level $Sp(n)$ factors to preserve the dimension of the Coulomb branch, and therefore determine special pieces. The map $d'$ satisfies this property in addition to $d' =d'^3$. The latter is required for it to describe the Higgs branch of $T_{\rho}(Sp(n)')$ theories discussed in \cite{Cremonesi:2014uva}.

We note there are many metaplectic special orbits which are not special, and vice versa. It would be nice to understand the relation of this map to the S-duality of boundary conditions in $\mathcal{N}=4$ super Yang-Mills theory\cite{Gaiotto:2008ak}.

\subsection{Constraints}
\subsubsection{\texorpdfstring{$(\pm i)$}{(±i)}-twisted sectors}
The local description of the Higgs field is as follows. Choose $X\in\mathfrak{j}_+$ be a nilpotent in a given metaplectic-special\footnote{We recall that a $C$-partition of $2n$ has odd parts occurring with even multiplicity. A metaplectic-special partition has the further property that there are an odd number of even parts between the beginning of the partition and any odd part. Thus $[4,3^2,2^2,1^2]$ is metaplectic-special, whereas $[6,3^2,2,1^2]$ is not.} ``Hitchin'' nilpotent orbit. $\mathcal{A}$ is the adjoint bundle on the punctured plane twisted by the outer-automorphism when we circle $z=0$. The Higgs field $\Phi(z)$ is a holomorphic section of $K\otimes\mathcal{A}$ with a \emph{pole} with residue $X$ at $z=0$. That is, it admits a Laurent expansion of the form
\begin{equation}
\Phi(z)= \frac{X}{z}+ \frac{A_i}{z^{3/4}}+ \frac{A_{-1}}{z^{1/2}}+ \frac{A_{-i}}{z^{1/4}}+\dots
\end{equation}
where the $A_\alpha$ are \emph{arbitrary} elements of $\mathfrak{j}_\alpha$. The characteristic polynomial
\[
\det(\lambda\Bid-\Phi(z))=\lambda^{2n+1}-\sum_{k=2}^{2n+1}\lambda^{2n+1-k}\phi_k(z)
\]
The $\phi_k(z)$ have Laurent expansions
\[
\begin{split}
\phi_{2l}(z)&=\frac{c^{(2l)}_{\pi_{2l}}}{z^{\pi_{2l}}}+\frac{c^{(2l)}_{\pi_{2l}-1}}{z^{\pi_{2l}-1}}+\dots+\frac{c^{(2l)}_{1}}{z}+...\\
\phi_{2l+1}(z)&=\frac{c^{(2l+1)}_{\pi_{2l+1}}}{z^{\pi_{2l+1}}}+\frac{c^{(2l+1)}_{\pi_{2l+1}-1}}{z^{\pi_{2l+1}-1}}+\dots+\frac{c^{(2l+1)}_{1/2}}{z^{1/2}}+...\\
\end{split}
\]
For $k$ even, $\pi_k\in \mathbb{N}$, whereas for $k$ odd $\pi_k\in \mathbb{N}+1/2$. The formula for the $\pi_k$ is as follows. Let $[P]$ be the partition of $(2n)$ corresponding to the Hitchin nilpotent orbit containing $X$.
\begin{itemize}
\item  Write $[P]$ as a Young diagram with columns given by the parts, $P_j$.
\item Number the boxes consecutively from $2$ to $(2n+1)$.
\item If $k$ occurs in the $j^{\text{th}}$ column, then $\pi_k=k-\chi_k$, where
\begin{equation}\label{chikdef}
\chi_k=\begin{cases}
\lfloor j/2\rfloor+1/2&k\; \text{odd}\\
\lceil j/2\rceil&k\; \text{even}
\end{cases}
\end{equation}
\end{itemize}

The Laurent coefficients, $c^{(k)}_l$ ($l>0$), are the local Hitchin base parameters. As in the untwisted type-D case \cite{Balasubramanian:2023iyx}, the \emph{leading} Laurent coefficients, $c^{(k)}_{\pi_k}$ can obey constraints\footnote{A proof of these constraints, along the lines of \cite{Balasubramanian:2023iyx} would use Spaltenstein's factorization \cite{spaltenstein1988polynomials} of the characteristic polynomial for $\mathfrak{sp}(2n)$ (just as \cite{Balasubramanian:2023iyx} used the factorization for $\mathfrak{so}(2n)$). But it would also require an analogue of the theorem proven in \cite{Distler:2024ckw}, which we currently do not possess and would have to use the embedding in $\mathfrak{sl}(2n+1)$. In lieu of that, we have checked that these constraints are satisfied, with the $a^{(k/2)}_{\pi_k/2}$  being explicit polynomials on the Lie algebra, for all of the metaplectic special orbits in $\mathfrak{j}_+\subset A_2,A_4,A_6,A_8$.}.
These come in two types.

Given a metaplectic-special Hitchin partition, $[P]$:
\begin{itemize}
\item[\bf Odd-type:] For every odd part (which necessarily occurs with even multiplicity), 
\[
P_{2j}=P_{2j+1}=\dots P_{2(j+l)-1}=2s-1
\]
let
\[
k=1+\sum_{i=1}^{2j-1} P_i
\]
Then, introducing a formal parameter $u$,  we have a constraint of the form
\[
u^{2l} c^{(k)}_{\pi_k} + u^{2l-1} c^{(k+2s-1)}_{\pi_k+1/2}+\dots+ c^{(k+2l(2s-1))}_{\pi_k+l}= 
\bigl(u^l a^{(k/2)}_{\pi_k/2}+u^{l-1}a^{(k/2+2s-1)}_{(\pi_k/2+1/4)}+\dots +a^{(k/2+l(2s-1)}_{(\pi_k+l)/2}\bigr)^2
\]
For instance, for Hitchin partition $[P]=[2,1^2]$, we have
\[
c^{(3)}_{5/2}= \bigl(a^{(3/2)}_{5/4}\bigr)^2,\qquad
c^{(4)}_{3}= 2a^{(3/2)}_{5/4}a^{(5/2)}_{7/4},\qquad
c^{(5)}_{7/2}= \bigl(a^{(5/2)}_{7/4}\bigr)^2
\]
\item[\bf Even-type:] For every ``marked pair" of even parts of the form
\[
P_{2j-1}=2r,\qquad P_{2j}= 2s,\qquad\text{with } r>s\geq0
\]
let
\[
k=1+\sum_{i=1}^{2j-1} P_i
\]
Then we have a constraint of the form
\begin{equation}\label{oddtype}
c_{\pi_k}^{(k)} = \left(a_{\pi_k/2}^{(k/2)}\right)^2
\end{equation}
For each marked pair, we have a choice of whether to impose the constraint \eqref{oddtype}. If there are $l$ marked pairs, there are $2^l$ possibilities. These are in 1-1 correspondence with the $2^l$ nilpotent orbits in the metaplectic-special piece on the Nahm side.
\end{itemize}

\subsubsection{Metaplectic Special Pieces and Metaplectic Sommers-Achar Groups}
The even-type constraints involve marked pairs
\[
P_{2j-1}=2r,\qquad P_{2j}= 2s,\qquad\text{with } r>s\geq 0
\]
in the Hitchin partition $[P]$. Label each marked pair by the corresponding positive integer $j$ and let
\[
S=\{j_1,j_2,\dots,j_l\},\qquad j_1< j_2<\dots<j_l
\]
be the ordered set of marked pairs. For each element $j\in S$, with $k_j=1+\sum_{i=1}^{2j-1}P_i$, we have a $\mathbb{Z}_2$ which acts as
\[
   a^{(k_j/2)}_{\pi_{k_j}/2}\to - a^{(k_j/2)}_{\pi_{k_j}/2}
\]
and for each $l$-element subset $S_i\subset S$, we have a metaplectic Sommers-Achar group, $C_i =\mathbb{Z}_2^l$ acting on the space of $a^{(k/2)}_{\pi_{k}/2}$s as above. The decision as to which of the constraints \eqref{oddtype} to impose is a choice of metaplectic Sommer-Achar group, $C_i$, to quotient by or, equivalently, to a choice of subset $S_i\subset S$.

On the Nahm side, we have a metaplectic special piece consisting of nilpotent orbits $[Q]$ which map to the same metaplectic special Hitchin orbit $d'([Q])=[P]$. One of these orbits is, itself, metaplectic special, $[Q]_s= d'([P])$. The rest are obtained as follows.

If $[Q]_s$ has a string of parts of the form (here, $m\geq 0$)
\[
\begin{split}
Q_{2(j-m)-1} &= 2r+2\\
Q_{2(j-m)}=Q_{2(j-m)+1}&=\dots=Q_{2j-1}=2r+1\\
Q_{2j}&=2r
\end{split}
\]
then we can do a metaplectic small degeneration (decreasing $Q_{2(j-m)-1}$ by 1 and increasing $Q_{2j}$ by 1)
\[
Q'_{2(j-m)-1} =
Q'_{2(j-m)}=Q'_{2(j-m)+1}=\dots=Q'_{2j-1}=Q'_{2j}=2r+1
\]
to form a new partition $[Q']$ which is necessarily metaplectic non-special. It is easy to see that $d'([Q'])=d'([Q]_s)=[P]$. So these are in the same metaplectic special piece.  Label the possible metaplectic small degeneration of $[Q]_s$ by the corresponding integers $j$. We obtain an ordered set
\[
S'=\{j_1,j_2,\dots,j_{l'}\},\qquad j_1<j_2<\dots<j_{l'}
\]
The Nahm orbits in the special piece containing $[Q]_s$ are in 1-1 correspondence with subsets $S'_i\subset S'$.

One can show\footnote{Using the methods that went into the proof of Theorem 3 in \cite{Balasubramanian:2023iyx}.} that $l'=l$. In fact, we can do better and give a \emph{formula} mapping small degenerations to marked pairs and vice versa. Let $[P]$ and $[Q]$ be dual metaplectic-special partitions
\[
d'([P])=[Q],\qquad d'([Q])=[P]
\]
Let
\[
\begin{split}
Q_{2(h-m)-1}&=2u+2\\
Q_{2(h-m)}&=Q_{2(h-m)+1}=\dots=Q_{2h-1}=2u+1\\
Q_{2h}&=2u
\end{split}
\]
be a metaplectic small degeneration in $[Q]$, which we label by the pair $(h,u)$, and let
\[
\begin{split}
P_{2j-1}&=2r\\
P_{2j}&=2s
\end{split}
\]
be a marked pair in $[P]$, which we label by the pair $(j,r)$. Then the map from small degenerations to marked pairs is given by
\begin{equation}
(j,r)= (u+1,h)
\end{equation}
Conversely, the map from marked pairs to metaplectic small degenerations is
\begin{equation}
(h,u)= (r,j-1)
\end{equation}

Let $\sigma: S'\to S$ be the order-reversing bijection between the ordered set of metaplectic small degenerations in the special Nahm partition $[Q]_s$ and the ordered set of marked pairs in the Hitchin partition, $[P]=d'([Q]_s)$. Then $\sigma$ induces a bijection between subsets $S'_i\subset S'$ and $S_i\subset S$. Hence it induces a bijection between Nahm orbits in the metaplectic special piece and metaplectic Sommers-Achar groups for the Hitchin partition. 
\[
[Q]_i \longleftrightarrow ([P],C_i)
\]
where $d'([Q]_i)=[P]$.

\subsection{The \texorpdfstring{$(-1)$}{(−1)}-twisted sector}
When we quotient by $\alpha^2$, the invariant subalgebra consists of matrices of the form
\[
\mathfrak{j}_+=\left\{\left.
\begin{pmatrix}
-tr(A)&0\\
0&A
\end{pmatrix}\right\vert A=\text{arbitrary} (2n)\times(2n)\; \text{matrix}\right\}=\mathfrak{sl}(2n)\oplus\mathfrak{u}(1)
\]
and the anti-invariant sector is
\[
\mathfrak{j}_-=\left\{\begin{pmatrix}
0&v^t\\
w&0
\end{pmatrix}\right\}=\begin{tikzpicture} 
 \draw[black,fill=white] (0.25, -0.25) rectangle (0.5, -0.5);
\end{tikzpicture} 
\oplus 
\overline{\begin{tikzpicture} 
 \draw[black,fill=white] (0.25, -0.25) rectangle (0.5, -0.5);
\end{tikzpicture} 
}
\]
The Higgs field locally takes the form
\[
\Phi(z)= \frac{X}{z}+\frac{A_-}{z^{1/2}} +\dots
\]
where $X$ is a nilpotent in some nilpotent orbit $[P]\subset\mathfrak{sl}(2n)\subset\mathfrak{j}_+$ and $A_-$ is an arbitrary element of $\mathfrak{j}_-$.
Since $\alpha^2$ is an inner automorphism of $\mathfrak{sl}(2n+1)$, the gauge-invariant information in $\Phi$ is the same as that contained in an untwisted (meromorphic) Higgs field with $X$ in a nilpotent orbit $[P']\subset \mathfrak{sl}(2n+1)$. The relation between $[P']$ and $[P]$ is to simply append a ``1" to the end of the partition: $[P']=[P,1]$.

Going to the Nahm side, to see the flavour symmetry associated to the $-1$ full puncture $[1^{2n}]$, note that we would expect to be able to glue two such punctures together via some exactly marginal gauging. To determine the gauge group, we can look at the Coulomb branch parameters added under such an operation. Relative to the original theory, removing the two $-1$ twisted punctures will remove $(2,4,6,\dots4n+2)$ from the graded Coulomb branch dimensions, where the number at position $i$ in the parenthesis indicate the number of CB parameters with scaling dimension $i+1$. There is also a global contribution to the graded CB dimensions given by  $(g-1)(3,5,7, \dots 4n+3)$, where $g$ is the genus of the Riemann surface. If there is a single theory that a handle is added to with the gluing, then the genus increases by one, resulting in a gain of $(3,5,7, \dots 4n+3)$. Alternatively, if we are gluing two separate Riemann surfaces, then the genus does not change, but there is still a gain of $(3,5,7, \dots 4n+3)$. Thus, in either case the change in the graded Coulomb branch dimensions is $(3,5,7, \dots 4n+3)-(2,4,6,\dots4n+2)=(1,1,\dots,1)$. We see a total of $2n$ Coulomb branch parameters are added from the gauge group with scaling dimensions $2,3,...,2n+1$ respectively. Hence, the gauging must be of an $SU(2n+1)$ factor, and thus any $-1$ twisted full puncture should contribute an $SU(2n+1)$ symmetry. The level for such a factor must be $k_{SU(2n+1)}=4n+2$, so that two can be glued together with vanishing $\beta$-function for the $SU(2n+1)$. But this is precisely the behaviour of the full puncture of the untwisted sector. So we expect to be able to identify the full puncture  of the $(-1)$-twisted sector with  the full puncture of the untwisted sector. Similarly, partial puncture closure leads us identify the rest of the punctures from the $(-1)$-twisted sector with a subset of the punctures from the untwisted sector; specifically, we map Hitchin partition $[P]$ to Hitchin partition $[P,1]$.

Under this identification, the fixtures from the $(i,i,-1)$ sector should be identical to (a subset of) the fixtures from the $(-i,i,1)$ sector.

%\subsection{Drop?}
%
%Let's work out the 
%details for the Hitchin system in the twisted $A_2$ case. The expansion of the Higgs field takes the form 
%\begin{equation*}
%    \Phi(z) = \frac{X}{z}+\frac{L}{z^{3/4}}+\frac{M}{z^{1/2}}+\frac{N}{z^{1/4}}+O+\dots
%\end{equation*}
%where 
%\begin{equation*}
%    X= \left(
%\begin{array}{ccc}
% 0 & 0 & 0\\
%  0 & 0 & 1 \\
% 0 & 0 & 0 \\
%\end{array}
%\right),
%\end{equation*}
%\begin{equation*}
%    L= \left(
%\begin{array}{ccc}
% 0 & l_1 & il_2\\
%  l_2 & 0 & 0 \\
% il_1 & 0 & 0 \\
%\end{array}
%\right),
%\end{equation*}
%\begin{equation*}
%    M= \left(
%\begin{array}{ccc}
% -2m & 0 & 0\\
%  0 & m & 0 \\
% 0 & 0 & m \\
%\end{array}
%\right),
%\end{equation*}
%\begin{equation*}
%    N= \left(
%\begin{array}{ccc}
% 0 & n_1 & -in_2\\
%  n_2 & 0 & 0 \\
% -in_1 & 0 & 0 \\
%\end{array}
%\right),
%\end{equation*}
%\begin{equation*}
%    O= \left(
%\begin{array}{ccc}
% 0 & 0 & 0\\
%  0 & o_1 & o_2 \\
% 0 & o_3 & -o_1 \\
%\end{array}
%\right),
%\end{equation*}
%Computing the characteristic polynomial gives the leading behaviors
%\begin{equation}
%    \phi_2 = \frac{3m^2+o_3+2l_1n_2}{z} +\dots
%\end{equation}
%and 
%\begin{equation}
%    \phi_3 = \frac{il_1^2}{z^{5/2}} + \dots
%\end{equation} 
%We see the numerator of the leading term is a perfect square and thus we have an $a$ constraint for $\phi_3$, in agreement with the prediction from \cite{Beem:2020pry}.

\section{Coulomb Branch Geometries}

Rather generally, the Coulomb branch geometry  of a 4d $\mathcal{N}=2$ theory is encoded in the data of a(n algebraic) complex integrable system. In class-S, this is invariably a Hitchin system, either on the punctured curve $C$ or (in the case of twisted theories) on a branched cover  $\tilde{C}\to C$, branched over the location of the insertions of twisted-sector punctures.

Here we will restrict ourselves to $C$ a 3-punctured sphere. Let $(x,y)$ be homogeneous coordinates on $\mathbb{CP}^1$. We will locate the $(\pm i)$-twisted punctures at $p_1=(1,0)$ and $p_2=(0,1)$ and the untwisted (or $(-1)$-twisted) puncture at $p_3=(1,1)$. Nominally, we want to pass to a 4-fold cover. But  everything will, in the end, factor through the double cover $x=\tilde{x}^2$, $y=\tilde{y}^2$.

Instead of letting $\Phi$ be a section of $K_C\otimes \mathcal{A}$ with poles at the $p_i$, we take it to be a \emph{holomorphic} section of $K_C(D)\otimes \mathcal{A}$, where $D=p_1+p_2+p_3$. Computing the characteristic polynomial,
\[
\det(\lambda\Bid -\Phi)=\lambda^{2n+1}-\sum_{k=2}^{2n+1} \lambda^{2n+1-k}\phi_k
\]
the $\phi_k$ are holomorphic sections of $K_C(D)^{\otimes k}$ with \emph{zeroes} of order $\chi_k=k-\pi_k$ at the punctures, where the $\chi_k$ are given by \eqref{chikdef}. In this convention, the spectral curve (Seiberg-Witten curve) is a compact curve, presented as a hypersurface in $K_C(D)=\mathcal{O}(1)$
and the Seiberg-Witten differential is
\begin{equation}\label{SWdef}
\lambda_{\text{SW}}= \frac{w(xdy-ydx)}{xy(x-y)}
\end{equation}
where $w$ is the fiber coordinate on $K_C(D)$. The spectral curve is presented as a polynomial in $(w,x^{1/2},y^{1/2})$ which is homogeneous (of bidegree $(2n+1,2n+1)$) under two $\mathbb{C}^*$ actions. Under the first, $w,x,y$ scale with weight-1 and the Coulomb branch parameters (which appear as coefficients in the polynomial) are invariant. Under the second, $x,y$ are invariant, $w$ scales with weight-1 and the Coulomb branch parameters $c_k$ scale with weight $k$. $\lambda_{\text{SW}}$ is invariant under the first $\mathbb{C}^*$ and scales with weight-1 under the second. Moreover, the coefficient of $w^k$ vanishes to order $\chi^{(i)}_k$ at $p_i$.

Pulled back to the double-cover, $\tilde{C}$,
\[
\lambda_{\text{SW}}= \frac{2w(\tilde{x}d\tilde{y}-\tilde{y}d\tilde{x})}{\tilde{x}\tilde{y}(\tilde{x}-\tilde{y})(\tilde{x}+\tilde{y})}
\]
In the formulas below, we won't always write explicitly the lift to $\tilde{C}$, but it should be understood.

\subsection{Examples}

\subsubsection{\texorpdfstring{$[1^2][1^2][1^3]$}{[1²][1²][1³]} , \texorpdfstring{$[1^2][2][1^3]$}{[1²][2][1³]} and \texorpdfstring{$[2][2][1^3]$}{[2][2][1³]}}

As a first example, take two full twisted punctures and one full untwisted puncture in the twisted $A_2$ theory. We list the zero-orders of the $\phi_k$ at each of the punctures in the table below

\begin{longtable}{|c|c|c|c|}
\hline
Nahm Orbit&Hitchin Orbit&$\chi_2$&$\chi_3$\\
\hline
\endhead
$[1^2]$&$([2],\mathbb{Z}_2)$&$1$&$1/2$\\
\hline
$[1^2]$&$([2],\mathbb{Z}_2)$&$1$&$1/2$\\
\hline
$[1^3]$&$[3]$&$1$&$1$\\
\hline
\end{longtable}
There are even-type constraints associated to each of the twisted punctures, which say that the coefficient of the leading term in the expansion of $\phi_3$ is a perfect square\footnote{Here, and in what follows, we slightly change notation and the subscripts denotes scaling dimension of the Coulomb branch parameters.} $c_3=a_{3/2}^2$. But $[1^2]$ is metaplectic non-special, so the metaplectic Sommers-Achar group is $\mathbb{Z}_2$ and we mod out by $a\to-a$, retaining $c$ as the Coulomb branch parameter.

The spectral curve is
\begin{equation}\label{223example}
\begin{split}
\Sigma =\bigl\{
0&=w^3-x^{1/2}y^{1/2}(x-y)(c_3 x-c'_3 y)\bigr)
\bigr\}
\end{split}
\end{equation}

If we replace the $[1^2]$ at $p_2$ by $[2]$ (which is metaplectic special), then we replace the corresponding Hitchin orbit by $([2],\Bid)$. Thus we don't mod out by $\mathbb{Z}_2$ and we should replace $c'_3$ by $a^{\prime\; 2}_{3/2}$ in \eqref{223example}. If we replace \emph{both} $[1^2]$ twisted punctures by $[2]$, then we don't mod out by either $\mathbb{Z}_2$ and the spectral curve reads
\begin{equation}\label{223thirdexample}
\begin{split}
\Sigma &=\bigl\{
0=w^3-x^{1/2}y^{1/2}(x-y)(a_{3/2}^2 x-a^{\prime\; 2}_{3/2} y\bigr)\bigr\}\\
&=\bigl\{
0=w^3 -\tilde{x}\tilde{y}(\tilde{x}-\tilde{y})(\tilde{x}+\tilde{y})(a_{3/2}\tilde{x}-a^{\prime}_{3/2}\tilde{y})(a_{3/2}\tilde{x}+a^{\prime}_{3/2}\tilde{y})
\bigr\}
\end{split}
\end{equation}
These three examples yield the Coulomb branch geometry for a trio of rank-2 theories with graded Coulomb branch dimensions $(d_{3/2},d_3)=(0,2)$, $(1,1)$ and $(2,0)$.

The last one is isomorphic to two copies of the rank-one $SU(3)$ instanton theory ($(A_1,D_4)$ Argyres-Douglas), which has a single Coulomb branch parameter of dimension 3/2. Presumably, one can see this directly from \eqref{223thirdexample} by a calculation similar to the one in \cite{Ergun:2020fnm}. The second one is the rank-two $SU(3)$ instanton theory. Both of these were identified in \cite{Beem:2020pry}.

\subsubsection{\texorpdfstring{$[4][1^4][3,2]$}{[4][1⁴][3,2]}}

\begin{longtable}{|c|c|c|c|c|c|}
\hline
Nahm Orbit&Hitchin Orbit&$\chi_2$&$\chi_3$&$\chi_4$&$\chi_5$\\
\hline
\endhead
$[4]$&$[2,1^2]$&$1$&$1/2$&$1$&$3/2$\\
\hline
$[1^4]$&$([4],\mathbb{Z}_2)$&$1$&$1/2$&$1$&$1/2$\\
\hline
$[3,2]$&$[2^2,1]$&$1$&$2$&$2$&$3$\\
\hline
\end{longtable}

This corresponds to the $D_{2}(SU(5))$ theory of \cite{Cecotti:2013lda,Cecotti:2012jx}. This time, there is an odd-type constraint associated to the puncture at $p_1$. This eliminates $c_3$, $c_4$ and $c_5$ in favour of $a_{3/2}$ and $a_{5/2}$:
\[
\begin{split}
c_{3}&=a_{3/2}^2\\
c_4&= 2a_{3/2}a_{5/2}\\
c_5&=a_{5/2}^2
\end{split}
\]
This yields the spectral curve
\begin{equation}\label{4111132example}
\begin{split}
\Sigma =\Bigl\{
0&=w^2\Bigl(w^3- x^{1/2}y^{1/2} (x-y)^2\bigl(w^2 a_{3/2}^2 +2 w x^{1/2} y^{1/2} a_{3/2}a_{5/2}+y(x-y)a_{5/2}^2\bigr)\Bigr)
\Bigr\}
\end{split}
\end{equation}
The trivial $w^l$ factor in this reducible  spectral curve is symptomatic of the presence of free hypers and, indeed, this fixure has two free hypers accompanying the rank-2 SCFT.

\subsubsection{\texorpdfstring{$[4,2^2][6,1^2][4^2,1]$}{[4,2²][6,1²][4²,1]}}

\begin{longtable}{|c|c|c|c|c|c|c|c|c|c|}
\hline
Nahm Orbit&Hitchin Orbit&$\chi_2$&$\chi_3$&$\chi_4$&$\chi_5$&$\chi_6$&$\chi_7$&$\chi_8$&$\chi_9$\\
\hline
\endhead
$[4,2^2]$&$([4,2^2],\Bid)$&$1$&$1/2$&$1$&$1/2$&$1$&$3/2$&$2$&$3/2$\\
\hline
$[6,1^2]$&$[4,1^4]$&$1$&$1/2$&$1$&$1/2$&$1$&$3/2$&$2$&$5/2$\\
\hline
$[4^2,1]$&$[3,2^2]$&$1$&$1$&$2$&$2$&$3$&$3$&$4$&$4$\\
\hline
\end{longtable}

$[4,2^2]$ is the metaplectic special orbit in a metaplectic special piece consisting of 4 orbits. There is a metaplectic small degeneration taking $[4,2^2]\to [3^2,2]$ and another taking $[4,2^2]\to [4,2,1^2]$. Applying both takes $[4,2]\to [3^2,1^2]$. All of these orbits map to the Hitchin orbit $[4,2^2]$ under $d'$. The latter has two marked pairs ($(4,2)$ and $(2,0)$), and hence two odd-type constraints
\[
\begin{split}
c_5&=a_{5/2}^2\\
c_9&= a_{9/2}^2
\end{split}
\]
The metaplectic special Nahm orbit $[4,2^2]$ corresponds to the trivial metaplectic Sommers-Achar group, which means we take $a_{5/2}$ and $a_{9/2}$ as Coulomb branch parameters. $[3^2,2]$ corresponds to modding out by
\[
a_{9/2}\to -a_{9/2}
\]
so that the Coulomb branch parameters are $a_{5/2}$ and $c_9$. $[4,2,1^2]$ corresponds to modding out by
\[
a_{5/2}\to -a_{5/2}
\]
so that the Coulomb branch parameters are $c_5$ and $a_{9/2}$. Finally, $[3^2,1^2]$ corresponds to modding out by the full $\mathbb{Z}_2\times \mathbb{Z}_2$, so that the Coulomb branch parameters are $c_5,c_9$.

The Nahm orbit $[6,1^2]$ maps to the Hitchin orbit $[4,1^4]$. The latter has an odd-type constraint, which sets
\[
\begin{split}
c_5&=a_{5/2}^2\\
c_6&=2a_{5/2}a_{7/2}\\
c_7&= a_{7/2}^2 +2 a_{5/2}a_{9/2}\\
c_8&=2a_{7/2}a_{9/2}\\
c_9&=a_{9/2}^2
\end{split}
\]
Placing $[4,2^2]$ at $p_1$, $[6,1^2]$ at $p_2$  and the untwisted puncture $[4^2,1]$ at $p_3$ leads to the spectral curve
\[
\begin{split}
\Sigma&=\Bigl\{
0=w^9 -w^6x^{1/2}y^{1/2}(x-y)(c_3x-c'_3y)-w^5xy(x-y)^2c_4-w^4x^{3/2}y^{3/2}(x-y)^2 c_5\\
&\qquad\qquad-w^3x^2y(x-y)^3 c_6-w^2 x^{5/2}y^{3/2}(x-y)^3 c_7\\
&\qquad\qquad -x^{1/2}y^{1/2}(x-y)^4\bigl(w^2 a_{5/2}+w x^{1/2}y^{1/2}a_{7/2} +x y a_{9/2}\bigr)^2\\
&\qquad\qquad-x^{5/2}y^{1/2}(x-y)^2 \bigl(
w^2 a'_{5/2}+iy(x-y)a'_{9/2}\bigr)\bigl(w^2 a'_{5/2}-iy(x-y)a'_{9/2}
\bigr)
\Bigr\}
\end{split}
\]
The coefficients on the 3rd line come from the odd-type constraints at $p_2$ and the coefficients on the 4th line come from the even-type constraints at $p_1$.

Replacing the Nahm orbit $[4,2^2]$ by one of the other orbits in the same metaplectic special piece affects only the coefficients last line. So, for instance, if we replace it by $[3^2,1^2]$, 
\[
\Sigma=\Bigl\{...-x^{5/2}y^{1/2}(x-y)^2\bigl(w^4c'_5+y^2(x-y)^2c'_9\bigr)\Bigr\}
\]

\subsubsection{\texorpdfstring{$[2n][2n][1^{2n+1}]$}{[2n][2n][1²ⁿ⁺¹]}}

As a final example, let us take two twisted simple punctures and one untwisted full puncture in the $A_{2n}$ theory. There are odd-type constraints at each of the twisted simple punctures, yielding the spectral curve
\[
\begin{split}
\Sigma=\Bigl\{
0=w^{2n+1}&-x^{1/2}y^{3/2}(x-y)\bigl(
w^{n-1}a_{3/2}+w^{n-2}x^{1/2}y^{1/2}a_{5/2}+\dots+x^{(n-1)/2}y^{(n-1)/2} a_{(2n+1)/2}
\bigr)^2\\
&-x^{3/2}y^{1/2}(x-y)\bigl(
w^{n-1}a'_{3/2}+w^{n-2}x^{1/2}y^{1/2}a'_{5/2}+\dots+x^{(n-1)/2}y^{(n-1)/2} a'_{(2n+1)/2}
\bigr)^2
\Bigr\}
\end{split}
\]
This is a product SCFT, consisting of two copies of the $D_{2}(SU(2n+1))$ theory, as identified in \cite{Beem:2020pry}

\section*{Acknowledgements}
 This work was supported in part by the National Science Foundation under Grant No.~PHY--2210562. 
\bibliographystyle{utphys}
%\small\baselineskip=.93\baselineskip
%\let\bbb\bibitem\def\bibitem{\itemsep1pt\bbb}
\bibliography{references}

\end{document}